\documentclass[runningheads]{llncs}
\usepackage[T1]{fontenc}
\usepackage{graphicx}
\usepackage{amsmath}
\usepackage{amssymb}
\usepackage{bm}
\usepackage{xcolor}

\begin{document}
\title{Convolutional autoencoder for the spatiotemporal latent representation of turbulence} %\thanks{Supported by organization x.}}
\titlerunning{CAE for latent representation of turbulence}
\author{Nguyen Anh Khoa Doan\inst{1} \and
Alberto Racca\inst{2,3} \and
Luca Magri\inst{3,4}}
\authorrunning{N. A. K. Doan et al.}
\institute{Delft University of Technology, Delft 2629HS, Netherlands \\
\email{n.a.k.doan@tudelft.nl} \and
University of Cambridge, Cambridge CB2 1PZ, United Kingdom \and
Imperial College London, London SW7 2AZ, United Kingdom \and
The Alan Turing Institute, London NW1 2DB, United Kingdom}
\maketitle              % typeset the header of the contribution
\begin{abstract}
% The abstract should briefly summarize the contents of the paper in
Turbulence is characterised by chaotic dynamics and a high-dimensional state space, which make this phenomenon challenging to predict. However, turbulent flows are often characterised by coherent spatiotemporal structures, such as vortices or large-scale modes, which can help obtain a latent description of turbulent flows. However, current approaches are often limited by either the need to use some form of thresholding on quantities defining the isosurfaces to which the flow structures are associated 
or the linearity of traditional modal flow decomposition approaches, such as those based on proper orthogonal decomposition.  
This problem is exacerbated in flows that exhibit extreme events, which are rare and sudden changes in a turbulent state.
The goal of this paper is to obtain an efficient and accurate reduced-order latent representation of a turbulent flow that exhibits extreme events. Specifically, we employ a three-dimensional multiscale convolutional autoencoder (CAE) to obtain such latent representation. We apply it to a three-dimensional turbulent flow.
We show that the Multiscale CAE is efficient, requiring  less than 10\% degrees of freedom than proper orthogonal decomposition for compressing the data and is able to accurately reconstruct flow states related to extreme events. The proposed deep learning architecture opens opportunities for nonlinear reduced-order modeling of turbulent flows from data.
\keywords{Chaotic System  \and Reduced Order Modelling \and Convolutional Autoencoder.}
\end{abstract}
\section{Introduction}

Turbulence is a chaotic phenomenon that arises from the nonlinear interactions between spatiotemporal structures over a wide range of scales. Turbulent flows are typically high-dimensional systems, which may exhibit sudden and unpredictable bursts of energy/dissipation \cite{Blonigan2019}. The combination of these dynamical properties makes the study of turbulent flows particularly challenging. Despite these complexities, advances have been made in the analysis of turbulent flows through the identification of coherent (spatial) structures, such as vortices \cite{Yao2020}, and modal decomposition techniques \cite{Berkooz1993}. These achievements showed that there exist energetic patterns within turbulence, which allow for the development of reduced-order models.
%can help us in  understanding turbulence and from which it may be possible to

To efficiently identify these patterns, recent works have used machine learning \cite{Brunton2020}. Specifically, Convolutional Neural Networks (CNNs) have been used to identify spatial features in flows \cite{Morimoto2021} and perform nonlinear modal decomposition \cite{Murata2019a,Fukami2020c}. These works showed the advantages of using CNNs over traditional methods based on Principal Component Analysis (PCA) (also called Proper Orthogonal Decomposition in the fluid mechanics community), providing lower reconstruction errors in two-dimensional flows. More recently, the use of such CNN-based architecture has also been extended to a small 3D turbulent channel flow at a moderate Reynolds number \cite{Nakamura2020}.
The works highlight the potential of deep learning for the analysis and reduced-order modelling of turbulent flows, but they were restricted to two-dimensional flows or weakly turbulent with non-extreme dynamics.  Therefore, the applicability of deep-learning-based techniques to obtain an accurate reduced-order representation of 3D flows with extreme events remains unknown. Specifically, the presence of extreme events is particularly challenging because they appear rarely in the  datasets.
In this paper, we propose a 3D Multiscale Convolutional Autoencoder (CAE) to obtain such a reduced representation of the 3D Minimal Flow Unit (MFU), which is a flow that exhibit such extreme events in the form of a sudden and rare intermittent quasi-laminar flow state. We  explore whether the Multiscale CAE is able to represent the flow in a latent space with a reduced number of degrees of freedom with higher accuracy than the traditional PCA, and whether it can also accurately reconstruct the flow state during the extreme events.

Section \ref{sec:MFU} describes the MFU and its extreme events. Section \ref{sec:Methodology} presents in detail the Multiscale CAE framework used to obtain a reduced-order representation of the MFU. The accuracy of the Multiscale CAE in reconstructing the MFU state is discussed in Section \ref{sec:Results}. A summary of the main results and directions for future work are provided in Section \ref{sec:Conclusion}.

\section{Minimal Flow Unit}
\label{sec:MFU}
The flow under consideration is the 3D MFU \cite{Jimenez1991}. The MFU is an example of prototypical near-wall turbulence, which consists of a turbulent channel flow whose dimensions are smaller than conventional channel flow simulations. % In this small domain, the flow accurately reproduces the near-wall turbulent statistics of turbulent channel flows. 
The system is governed by the incompressible Navier-Stokes equations %, with initial conditions 
\begin{align}
\begin{split}
    \nabla \cdot \bm{u} & = 0, \\
    \partial_t \bm{u} + \bm{u} \cdot \nabla \bm{u} & = \frac{1}{\rho} \bm{f_0} - \frac{1}{\rho} \nabla p + \nu \Delta \bm{u}, 
\end{split}
\end{align}
where $\bm{u}=(u,v,w)$ is the 3D velocity field and 
$\bm{f}_0 = (f_0,0,0)$ is the constant forcing in the streamwise direction, $x$; $\rho$, $p$, and $\nu$ are the density, pressure, and kinematic viscosity, respectively.
%The geometrical domain of the MFU is composed of two walls,
In the wall-normal direction, $y$,  we impose a no-slip boundary condition, $\bm{u}(x,\pm \delta,z,t) = 0$, where $\delta$ is half the channel width. In the streamwise, $x$, and spanwise, $z$, directions we have periodic boundary conditions. 
For this study, a channel with dimension $\Omega\equiv \pi \delta \times 2\delta \times 0.34\pi \delta$ is considered, as in  \cite{Blonigan2019} with $\delta=1.0$. The Reynolds number of the flow, which is based on the bulk velocity and the half-channel width, is set to $Re=3000$, which  corresponds to a friction Reynolds number $Re_\tau \approx 140$.
An in-house code similar to that of \cite{Bernardini2014} is used to simulate the MFU, and generate the dataset on which the Multiscale CAE is trained and assessed.
% 
% We employ an in-house code based on work by \cite{Bernardini2014} to simulate the MFU. The flow is discretized on a Cartesian grid with staggered central second-order finite-difference approximations. Time marching is performed with a third-order low-storage Runge-Kutta algorithm coupled with a second-order Crank-Nicolson scheme, which are combined in a fractional-step procedure in which the convective and diffusive terms are treated explicitly and implicitly, respectively \cite{Bernardini2014}. The solution of the Poisson equation for the pressure is obtained with a direct solver based on the Fourier transform \cite{Kim1985}.

% \subsection{Extreme events}

\begin{figure}
    \centering
    \includegraphics[width=\textwidth]{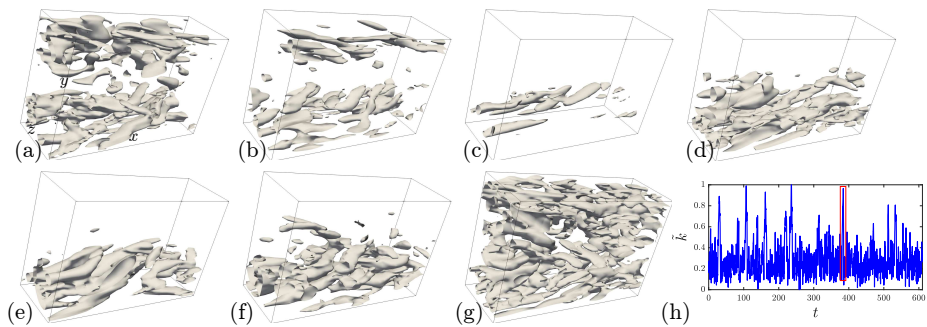}
    \caption{(a-g) Snapshots of the $Q$-criterion isosurface (with value $Q=0.1$) during an extreme event, where $Q=0.5(||\bm{\omega}||^2 - ||\bm{S}||^2)$,  $\bm{\omega}$ is the vorticity vector, and $\bm{S}$ is the strain-rate tensor. (h) Evolution of kinetic energy, $k$, of the MFU. The red box indicates the event whose evolution is shown in (a-g).}
    \label{fig:MFU}
\end{figure}

The extreme events in the MFU are quasi-relaminarization events, which take place close to either wall. A typical evolution of the flow during a quasi-relaminarization event is shown in Fig. \ref{fig:MFU}(a-g). Time is normalized by the eddy turnover time.
During an extreme event, 
(i) the flow at either wall (the upper wall in Fig. \ref{fig:MFU}) becomes laminar (Fig. \ref{fig:MFU}(a-c));
(ii) the flow remains laminar for some time (Fig. \ref{fig:MFU}(c-f)), which results in a larger axial velocity close to the centerline (and therefore an increase in kinetic energy);
(iii) the greater velocity close to the centerline makes the effective Reynolds number of the flow larger, which in turn makes the flow prone to a turbulence burst on the quasi-laminar wall;
(iv) the turbulence burst occurs on the quasi-laminar wall, which results in a large increase in the energy dissipation rate;
and (v) the flow close to that quasi-laminar wall becomes turbulent again, which leads to a decrease in the kinetic energy (Fig. \ref{fig:MFU}(g)). % \lm{[If you need space, I would shortened the previous thorough physical description.]}

These quasi-relaminarisation are accompanied by bursts in the total kinetic energy, $k(t)=\int\int\int_\Omega \frac{1}{2} \bm{u}\cdot \bm{u} dx dy dz$, where $\Omega$ is the computational domain. This can be seen in Fig. \ref{fig:MFU}h, where the normalized kinetic energy, $\tilde{k}=(k-\min(k))/(\max(k)-\min(k))$ is shown.

% , and its dissipation rate, $D(t)$. These are computed as
% \begin{equation}
%     , \hspace{11pt}
%     D(t) = \int\int\int_\Omega \textrm{tr}\left(\bm{\tau} \nabla \bm{u} \right) dx dy dz,
% \end{equation}
% where $\Omega$ is the computational domain and  $\bm{\tau} = \mu \left(\nabla \bm{u} + \nabla \bm{u}^T \right)$ is the stress tensor, with $\mu$ being the dynamic viscosity. 
% \begin{figure}
%     \centering
%     \includegraphics[width=0.75\textwidth]{Fig_1_MFU_k_D_evol.eps}
%     \caption{Time evolution of the kinetic energy (a) and  dissipation rate (b) in the minimal flow unit. Extreme events correspond to the peaks in $k$ and $D$.}
%     \label{fig:evol_k_Z}
% \end{figure}

% Figure \ref{fig:evol_k_Z} shows the time evolution of $k$ and $D$,

The dataset of the MFU contains 2000 eddy turnover times (i.e., 20000 snapshots) on a grid of $32\times 256 \times 16$, which contains 50 extreme events. The first 200 eddy turnover times of the dataset (2000 snapshots) are employed for the training of the Multiscale CAE, which contains only 4 extreme events.

% Because an extreme event is localized at either wall, the total kinetic energy computed over the whole domain makes the identification of specific extreme events difficult, and only very strong extreme events where the flow near both walls simultaneously becomes quasi-laminar are apparent in the time series of total kinetic energy. Those are the peaks observed in Figure \ref{fig:evol_k_Z}. 
% Therefore, we define a near-wall kinetic energy deficit to identify extreme events. This is defined as the deficit of kinetic energy in a layer at either wall, which occurs when it becomes quasi-laminar. Hence, we first define  
% \begin{equation}
%     k^u = \int^{\delta}_{0.75\delta}\int\int \frac{1}{2} \bm{u}\cdot \bm{u} dx dz dy, \hspace{11pt} k^l =  \int^{-0.75\delta}_{-\delta}\int\int \frac{1}{2} \bm{u}\cdot \bm{u} dx dz dy
% \end{equation}
% where $k^u$ and $k^l$ are the near-wall kinetic energy of the flow at the upper and lower wall, respectively, and compute the normalized kinetic energy deficit as $\widetilde{k}_w^\cdot = 1- k_w^\cdot / \max(k_w^\cdot)$ where $\cdot$ is $l$ ($u$) for the lower (upper) wall. The symbol $\tilde{\cdot}$ indicates the normalization. 
% We define an extreme event as the event peak that occurs when either $k_w^u$ or $k_w^l$ are greater than $0.55$. This threshold was chosen to ensure that only clear peaks of kinetic energy are considered as extreme events. Such a value indeed indicates a large deficit in kinetic energy near that wall, and thus a quasi-laminar flow state.

\section{Multiscale Convolutional Autoencoder}
\label{sec:Methodology}
\begin{figure}
    \centering
    \includegraphics[width=0.9\textwidth]{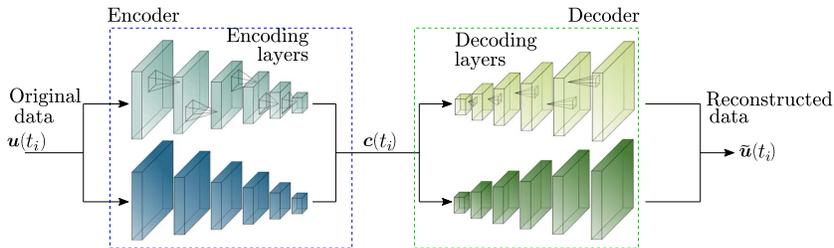}
    \caption{Schematic of the 3D multiscale convolutional autoencoder.}
    \label{fig:MSAE}
\end{figure}

We implement a 3D convolutional autoencoder. It should be noted that we only consider CNN-based autoencoder here and not other approaches such as transformer-based ones (like in \cite{Dosovitskiy2021}) as the latter have not yet shown to be widely applicable to dataset with a strong spatial based information (such as images or flow dataset). A schematic of the proposed architecture is shown in Fig. \ref{fig:MSAE}. 

%\subsection{Convolutional autoencoder}
%\label{sec:AE}
The three dimensional convolution autoencoder (CAE) learns an efficient reduced-order representation of the original data, which consists of the flow state $\bm{u}\in \mathbb{R}^{N_x\times N_y \times N_z \times N_u}$, where $N_x$, $N_y$ and $N_z$ are the number of grid points
, and $N_u=3$ is the number of velocity components. On one hand, the encoder (blue box in Fig. \ref{fig:MSAE}) reduces the dimension of the data down to a latent state, $\bm{c}$, with  small dimension $N_c \ll N_xN_yN_zN_u$. This operation can be symbolically expressed as $\bm{c}=\mathcal{E}(\bm{u};\bm{\phi}_E)$, where $\bm{\phi}_E$ represents the weights of the encoder. On the other hand, the decoder (green box in Fig. \ref{fig:MSAE}) reconstructs the data from the latent state back to the original full flow state. This  operation is expressed as $\widetilde{\bm{u}}=\mathcal{D}(\bm{c};\bm{\phi}_D)$, where $\bm{\phi}_D$ are the trainable weights of the decoder.
We employ a multiscale autoencoder, which was originally developed for image-based super-resolution analysis \cite{Du2018}. We use here a multiscale autoencoder and not a standard one as previous works \cite{Fukami2019,Hasegawa2020} have demonstrated the ability of the multiscale version in leveraging the multiscale information in turbulent flows to better reconstruct the flow from the latent space. It relies on the use of convolutional kernels of different sizes to analyse the input and improve reconstruction in fluids \cite{Hasegawa2020,Racca2022a}.
In this work, two kernels, $(3\times 5 \times 3)$ and $(5 \times 7 \times 5)$, are employed (represented schematically by the two parallel streams of encoder/decoder in the blue and green boxes in Fig. \ref{fig:MSAE}). This choice ensures a trade-off between the size of the 3D multiscale autoencoder and the reconstruction accuracy (see Section~\ref{sec:Results}).
To reduce the dimension of the input, the convolution operation in each layer of the encoder is applied in a strided manner, which means that the convolutional neural network (CNN) kernel is progressively applied to the input by moving the CNN kernel by $(s_x,s_y,s_z)=(2,4,2)$ grid points. This results in an output of a smaller dimension than the input. After each convolution layer, to fulfill the boundary conditions of the MFU, periodic padding is applied in the $x$ and $z$ directions, while zero padding is applied in the $y$ direction.
Three successive layers of CNN/padding operations are applied to decrease the dimension of the original field from $(32,256,16,3)$  to $(2,4,2,N_f)$, where $N_f$ is the specified number of filters in the last encoding layer. As a result, the dimension of the latent space is $N_c =16\times N_f$.
The decoder mirrors the architecture of the encoder, where transpose CNN layers \cite{Zeiler2010} are used, which increase the dimension of the latent space up to the original flow dimension. The end-to-end autoencoder is trained by minimizing the mean squared error (MSE) between the reconstructed velocity field, $\widetilde{\bm{u}}$, and the original field, $\bm{u}$ using the ADAM optimizer.

\section{Reconstruction error}
\label{sec:Results}

\begin{figure}
    \centering
    \includegraphics[width=0.9\textwidth]{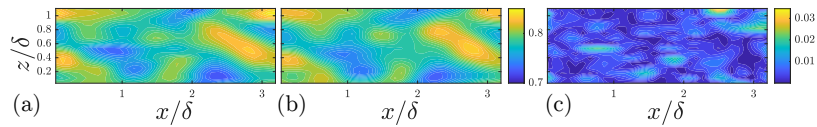}
    \caption{Comparison of (a) the actual velocity magnitude (ground truth), (b) the CAE-reconstructed velocity magnitude, (c) the root-squared difference between (a) and (b) in the mid-$y$ plane for a typical snapshot in the test set.}
    \label{fig:AE_recons_comp}
\end{figure}

We analyze the ability of the CAE to learn a latent space that encodes the flow state accurately. To do so, we train three CAEs with latent space dimensions of $N_c=384$, $768$, and $1536$. The training of each CAE took between 8 and 16 hours using 4 Nvidia V100S (shorter training time for the model with the smaller latent space). After training, the CAE can process 100 samples in 1.8s to 5s depending on the dimension of the latent space (the faster processing time corresponding to the CAE with the smaller latent space). We compute their reconstruction errors on the test set based on the MSE. A typical comparison between a reconstructed velocity field obtained from the CAE with $N_c=1536$ is shown in Fig. \ref{fig:AE_recons_comp}. The CAE is able to reconstruct accurately the features of the velocity field. 

\begin{figure}
    \centering
    \includegraphics[width=0.55\textwidth]{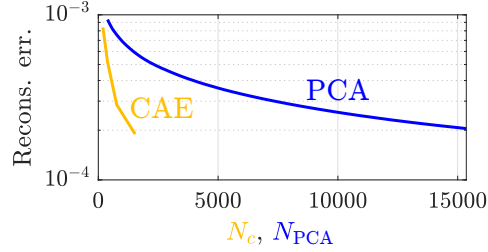}
    \caption{Reconstruction error with a PCA-based method (blue) and the autoencoder (yellow) for different dimensions of the latent space, $N_c$, or number of retained PCA components, $N_{PCA}$. The reconstruction error is computed as the mean squared error between the reconstructed velocity field and the exact field, averaged over the test set. }
    \label{fig:AE_MSE_modes}
\end{figure}

To provide a comparison with the CAE, we also compute the reconstruction error obtained from PCA, whose principal directions are obtained with the method of snapshots \cite{Berkooz1993} on the same dataset. 
Figure \ref{fig:AE_MSE_modes} shows the reconstruction error. 
PCA decomposition requires more than 15000 PCA components to reach the same level of accuracy as the CAE with a latent space of dimension 1536. This highlights the advantage of learning a latent representation with nonlinear operations, as in the autoencoder, compared with relying on a linear combination of components, as in PCA.

\begin{figure}
    \centering
    \includegraphics[width=0.55\textwidth]{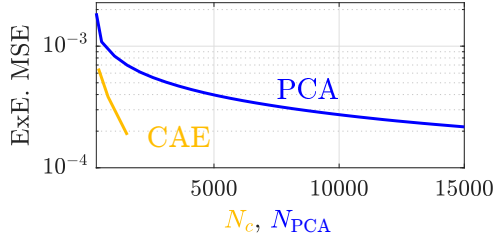}
    \caption{Reconstruction error on the extreme events flow states with PCA-based method (blue) and autoencoder (yellow) for different dimensions of the latent space, $N_c$, or number of retained PCA components, $N_{PCA}$. The reconstruction error is computed as in Fig. \ref{fig:AE_MSE_modes} but only on the flow state corresponding to extreme events.}
    \label{fig:AE_ExE_MSE_modes}
\end{figure}

\begin{figure}
    \centering
    \includegraphics[width=0.8\textwidth]{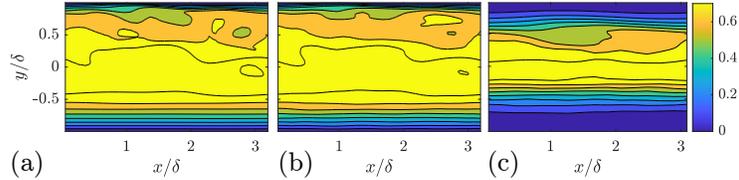}
    \caption{Comparison of (a) the velocity magnitude (ground truth), (b) the CAE-reconstructed velocity magnitude, (c) the PCA-reconstructed velocity magnitude in the mid-$z$ plane for a typical extreme event snapshot in the test set.}
    \label{fig:AE_ExE_MSE_modes_fields}
\end{figure}

The better performance of the CAE with respect to PCA is evident when we consider the reconstruction accuracy for flow states that correspond to extreme events, which we extract 
from the test set. Here, we define an extreme event as a flow state with normalized kinetic energy above a user-selected threshold value of 0.7. 
Hence, for the selected snapshots in the test set, the mean squared error (MSE) between the reconstructed velocity and the truth is computed using the CAE and PCA as a function of the latent space size. The resulting MSE is shown in Fig. \ref{fig:AE_ExE_MSE_modes}, where the CAE exhibits an accuracy for the extreme dynamics similar to the reference case of the entire test set (see Fig. \ref{fig:AE_MSE_modes}). On the other hand, the accuracy of the PCA is lower than the reference case of the entire dataset. This lack of accuracy is further analysed in Fig. \ref{fig:AE_ExE_MSE_modes_fields}, where typical velocity magnitudes, i.e. the norm of the velocity field $\bm{u}$, 
are shown for the mid-$z$ plane during a representative extreme event.
The velocity reconstructed with the CAE (Fig. \ref{fig:AE_ExE_MSE_modes_fields}b) is almost identical to the true velocity field (Fig. \ref{fig:AE_ExE_MSE_modes_fields}a). The CAE captures the smooth variation at the lower wall indicating a quasi-laminar flow state in that region. In contrast, the velocity field reconstructed using PCA (Fig. \ref{fig:AE_ExE_MSE_modes_fields}c) completely fails at reproducing those features. This is because extreme states are rare in the training set used to construct the PCA components (and the CAE). Because of this, the extreme states only have a small contribution to the PCA components and are accounted for only in higher order PCA components, which are neglected in the 1536-dimensional latent space. 

This result indicates that only the CAE and its latent space can be used for further applications that requires a reduced-order representation of the flow, such as when trying to develop a reduced-order model. This is also supported by findings in \cite{Racca2022a} where it was shown that a similar CAE could be used in combination with a reservoir computer to accurately forecast the evolution of a two dimensional flow.

\section{Conclusion}
\label{sec:Conclusion}

In this work, we develop a nonlinear autoencoder to obtain an accurate latent representation of a turbulent flow that exhibits extreme events. We propose the 3D Multiscale CAE to learn the spatial features of the MFU, which exhibits extreme events in the form of near-wall quasi-relaminarization events. The model consists of a convolutional autoencoder with multiple channels, which learn an efficient reduced latent representation of the flow state. We apply the framework to a three-dimensional turbulent flow with extreme events (MFU). We show that the Multiscale CAE is able to compress the flow state  to a lower-dimensional latent space by three orders of magnitude to accurately reconstruct the flow state from this latent space. This constitutes a key improvement over principal component analysis (PCA), which requires at least one order of magnitude more PCA components to achieve an accuracy similar to the CAE. This improvement in reconstruction accuracy is crucial for the reconstruction of the flow state during the extreme events of the MFU. This is because extreme states are rare and, thus, require a large number of PCA components to be accurately reconstructed.

The proposed method and results open up possibilities for using deep learning to obtain an accurate latent reduced representation of 3D turbulent flows. Future work will be devoted to physically interpreting the latent space discovered by the Multiscale CAE, and learning the dynamics in this latent space.

\subsubsection{Acknowledgements} 
The authors thank Dr. Modesti for providing the flow solver. N.A.K.D and L.M acknowledge that part of this work was performed during the 2022 Stanford University CTR Summer Program.
L.M. acknowledges the financial support from the ERC Starting Grant PhyCo 949388. A. R. is supported by the Eric and Wendy Schmidt AI in Science Postdoctoral Fellowship, a Schmidt Futures program. 

\bibliographystyle{splncs04}
\bibliography{library2}
\end{document}